\documentstyle[12pt]{article}

\catcode`\@=11
\long\def\@makefntext#1{
\protect\noindent \hbox to 3.2pt {\hskip-.9pt
$^{{\ninerm\@thefnmark}}$\hfil}#1\hfill}		

\def\@makefnmark{\hbox to 0pt{$^{\@thefnmark}$\hss}}  

\def\ps@myheadings{\let\@mkboth\@gobbletwo
\def\@oddhead{\hbox{}
\rightmark\hfil\ninerm\thepage}
\def\@oddfoot{}\def\@evenhead{\ninerm\thepage\hfil
\leftmark\hbox{}}\def\@evenfoot{}
\def\sectionmark##1{}\def\subsectionmark##1{}}
\renewcommand{\thefootnote}{\fnsymbol{footnote}}

\newcounter{appendixc}
\newcounter{subappendixc}[appendixc]
\newcounter{subsubappendixc}[subappendixc]

\renewcommand{\appendix}[1] {\vspace*{0.6cm}
        \refstepcounter{appendixc}
        \setcounter{figure}{0}
        \setcounter{table}{0}
        \setcounter{equation}{0}
        \renewcommand{\thefigure}{\Alph{appendixc}.\arabic{figure}}
        \renewcommand{\thetable}{\Alph{appendixc}.\arabic{table}}
        \renewcommand{\theappendixc}{\Alph{appendixc}}
        \renewcommand{\theequation}{\Alph{appendixc}.\arabic{equation}}
        \noindent{\bf Appendix \theappendixc #1}\par\vspace*{0.4cm}}

\renewenvironment{thebibliography}[1]
    {\begin{list}{\arabic{enumi}.}
    {\usecounter{enumi}\setlength{\parsep}{0pt}
\setlength{\leftmargin 1.25cm}{\rightmargin 0pt}
     \setlength{\itemsep}{0pt} \settowidth
    {\labelwidth}{#1.}\sloppy}}{\end{list}}

\newcounter{itemlistc}
\newcounter{romanlistc}
\newcounter{alphlistc}
\newcounter{arabiclistc}

\newcommand{\fcaption}[1]{
        \refstepcounter{figure}
        \setbox\@tempboxa = \hbox{\footnotesize Fig.~\thefigure. #1}
        \ifdim \wd\@tempboxa > 6in
           {\begin{center}
        \parbox{6in}{\footnotesize\baselineskip=15pt Fig.~\thefigure. #1}
            \end{center}}
        \else
             {\begin{center}
             {\footnotesize Fig.~\thefigure. #1}
              \end{center}}
        \fi}
\newcommand{\tcaption}[1]{
        \refstepcounter{table}
        \setbox\@tempboxa = \hbox{\footnotesize Table~\thetable. #1}
        \ifdim \wd\@tempboxa > 6in
           {\begin{center}
        \parbox{6in}{\footnotesize\baselineskip=15pt Table~\thetable. #1}
            \end{center}}
        \else
             {\begin{center}
             {\footnotesize Table~\thetable. #1}
              \end{center}}
        \fi}

 1 
 1  
 1  
 1  
 1  
 1  
 1  

\font\ninerm=cmr9

\evensidemargin=\oddsidemargin
\input{psfig.tex}

\begin{document}


July, 1995 \hfill hep-ph/9507454
\begin{flushright}
VPI-IHEP-95-12\\
TUIMP-TH-95/67\\
MSUHEP-50712
\end{flushright}
\vskip 0.4in
\centerline{\normalsize\bf
Longitudinal/Goldstone Boson Equivalence and Phenomenology\\}
\baselineskip=15pt
\centerline{\normalsize\bf
of Probing the Electroweak Symmetry Breaking
\footnote{Talk presented by Yu-Ping Kuang at First International Conference
on Frontiers of Physics,\\
\hspace{0.6cm}Aug.5-9, 1995, Shantou, China.}}

\vspace*{0.7cm}
\centerline{\normalsize {\bf  Hong-Jian He} }
\baselineskip=13pt
\centerline{\normalsize\it
Department of Physics and Institute of High Energy Physics}
\baselineskip=13pt
\centerline{\normalsize\it
Verginia Polytechnic Institute and State University}
\baselineskip=14pt
\centerline{\normalsize\it Blacksburg, Virginia 24061-0435, USA}
\vspace*{0.2cm}
\centerline{\normalsize\bf Yu-Ping Kuang}
\baselineskip=13pt
\centerline{\normalsize\it
Institute of Modern Physics, Tsinghua University, Beijing 100084, China}

\vspace*{0.2cm}
\centerline{\normalsize\bf C.--P. Yuan}
\baselineskip=13pt
\centerline{\normalsize\it
Department of Physics and Astronomy, Michigan State University }
\baselineskip=13pt
\centerline{\normalsize\it East Lansing, Michigan 48824, USA}

\baselineskip=13pt
\vspace{0.6cm}
\begin{abstract}

\baselineskip=13pt
\noindent
We formulate the Equivalence between the longitudinal weak-boson and
the Goldstone boson as a criterion for sensitively
probing the electroweak symmetry
breaking mechanism and develop a precise power
counting rule for chiral Lagrangian formulated electroweak theories.
With these we semi-quantitatively analyze the sensitivities to
various effective operators related to electroweak symmetry breaking
via weak-boson scatterings at the CERN Large Hadron Collider (LHC).
\end{abstract}

\baselineskip=15pt
\setcounter{footnote}{00}
\renewcommand{\thefootnote}{\alph{footnote}}
\renewcommand{\baselinestretch}{1.3}

\vspace{0.6cm}

Recent LEP/SLC experiments can test the electroweak (EW) theory to the
accuracy of one-loop corrections, and support the spontaneously broken
$~SU(2)\times U(1)~$ gauge theory as
the correct theory of the EW interactions.
However, light Higgs boson has not been
found, and the current experiments are
insensitive to the spontaneous symmetry
breaking (SSB) sector of the theory,
compatible with a wide
range of the Higgs boson mass $60$GeV$\leq m_H\leq 1$TeV. So the
SSB mechanism in the EW
theory is still a mystery, and it is thus important to
probe {\it all possible} SSB
mechanisms: either weakly or strongly interacting.

We know that only the longitudinal component $V^a_L$
of the weak-boson $V^a$
($W^\pm$,$Z^0$) (arising from ``eating'' the
would-be Goldstone boson (GB)) is
sensitive to the SSB sector, while
the transverse component $V^a_T$ is
not. The physical $V^a_L$ scattering amplitude
is quantitatively related to
the corresponding GB amplitude by
the electroweak Equivalence Theorem (ET)$^{1\sim 3}$ which
comes from the following ET identity$^{2\sim 3}$
$$
T[V^{a_1}_L,\cdots ,V^{a_n}_L;\Phi_{\alpha}]
= C\cdot T[-i\pi^{a_1},\cdots ,-i\pi^{a_n};\Phi_{\alpha}]+ B  ~~,
\eqno(1)                                               
$$
$$
\begin{array}{ll}
C & \equiv C^{a_1}_{mod}\cdots C^{a_n}_{mod} ~, \\
B & \equiv\sum_{l=1}^n (~C^{a_{l+1}}_{mod}\cdots C^{a_n}_{mod}
T[v^{a_1},\cdots ,v^{a_l},-i\pi^{a_{l+1}},\cdots ,
-i\pi^{a_n};\Phi_{\alpha}]\\
  & ~~~~~~~~~~~~~~+ ~{\rm permutations ~of}~v'
{\rm s ~and}~\pi '{\rm s}~)~,
  \\[0.2cm]
v^a & \equiv v^{\mu}V^a_{\mu} ~,
{}~~~~v^{\mu}~\equiv \epsilon^{\mu}_L-k^\mu /M_a
= O(M_a/E) ~,~~~(M_a=M_W,M_Z)~,
\end{array}
\eqno(2)                                            
$$
where $~\pi^a$'s  are GB fields,
and $\Phi_{\alpha}$ denotes other possible
physical in/out states. The renormalization scheme-dependent constant
modification factor $~C_{mod}^a~$ has been generally studied in Ref.2-3,
which can be exactly simplified as $~C^a_{mod}=1~$ in certain convenient
renormalization schemes$^{3\sim 4}$.

For strongly interacting SSB models,
the $V^a_L$-amplitude on the
L.H.S. of
(1) is experimentally measurable, while the GB-amplitude on the R.H.
S. of (1), though not directly measurable, carries
the information about the
SSB mechanism. Similar to $V^a_T$, the $B$-term
in (1) is not sensitive to the
SSB mechanism. If, under certain conditions, the
$B$-term can be neglected,
(1) reveals the {\it equivalence} between the $V^a_L$-amplitude
and the GB-amplitude. In this case the $V^a_L$-scattering experiments
can be used to sensitively and unambiguously probe
the SSB mechanism. When $B$ is not
negligible, measurements of the $V^a_L$, $V^a_T$ and $B$
amplitudes with higher precision will be required for
probing the SSB mechanism, and those
expreriments at LHC will be harder.

The conditions for neglecting the $B$-term
in (1), i.e. the condition for the
validity of the ET, is
actually subtle. We first note that the spin-$0$ GB's
are invariant under the proper Lorentz transformations,
while, on the contrary,
$V_L$, $V_T$ and $B$ are
Lorentz non-invariant. Therefore the ratio of the
$B$-magnitude relative to the GB-amplitude
in (1) is Lorentz frame dependent.
So neglecting $B$ makes sense
only if the Lorentz frame belongs to a group of
frames within which Lorentz
transformation does not significantly enhance
$B$. We call such frames {\it safe frames}.
The condition for a Lorentz
frame to be {\it safe} is given in Ref.3, which is
$$
E_j\sim k_j \gg M_W , \hspace{2.0cm}  (~j=1,2,\cdots,n~)~,~
$$
where $E_j$ is the energy
of the $j$-th external $V^a_L$-line. For a given
process, $E_j$ can be easily
obtained from the kinematics. So this condition is a
convenient criterion for judging whether
the experimental center of mass
frame is {\it safe} or
not for a given process, i.e. it can discriminate
processes which are not sensitive
for probing the SSB mechanism\footnote{See
the example given in Ref.3.}. With this consideration, the ET can be
precisely formulated as$^3$
$$
T[V^{a_1}_L,\cdots ,V^{a_n}_L;\Phi_{\alpha}]
= C\cdot T[-i\pi^{a_1},\cdots ,-i\pi^{a_n};\Phi_{\alpha}]+
O(M_W/E_j{\rm -suppressed} ),
\eqno(3a)                                          
$$
$$
\begin{array}{l}
E_j \sim k_j  \gg  M_W , ~~~~~(~ j=1,2,\cdots ,n ~)~~;
\end{array}
\eqno(3b)                                         
$$
$$
\begin{array}{l}
B  \ll  C\cdot T[-i\pi^{a_1},\cdots ,-i\pi^{a_n};\Phi_{\alpha}] ~~.
\end{array}
\eqno(3c)                                         
$$
(3b) and (3c) are the conditions for neglecting the $B$-term in (1)
(for the validity of the ET), or
the conditions for sensitively probing the
SSB mechanism via $V^a_L$-scattering experiments. Here we see the
{\it profound physical content of
the ET}, i.e. ET is not merely a tool for
simplifying calculations.

 The next thing is to realize the
quantitative meaning of the condition (3c).
To a given order $N$ in a perturbative expansion,
the amplitude $T$ can be
written as $T=\sum_{\ell=0}^NT_\ell$ with $T_0>T_1,\cdots,T_N$.
Let $T_{\min}=\{T_0,\cdots,T_N \}_{\min}$. Then, to the precision
of $T_{\min}$, condition (3c) precisely implies$^3$
$$
\begin{array}{ll}
B & \approx O(\frac{M_W^2}{E_j^2}) \,T_0[ -i\pi^{a_1},\cdots , -i\pi^{a_n};
\Phi_{\alpha}] +
  O(\frac{M_W}{E_j}) \,T_0[ V_{T_j}^{a_{r_1}}, -i\pi^{a_{r_2}},
                      \cdots , -i\pi^{a_{r_n}}; \Phi_{\alpha}] \\
 & \ll  T_{\min}[-i\pi^{a_1},\cdots ,-i\pi^{a_n};\Phi_{\alpha}]  ~~.
\end{array}
\eqno(4)                                                 
$$

In the chiral Lagrangian formulated EW
theory (CLEWT), the $~O(E^2)~$ leading
amplitude $T_0$ is model-independent. Thus,
for probing the SSB mechanism, we
should take into account the next-to-leading
$~O(E^4)~$ amplitude $T_1$, i.e.
$~T_{\min}=T_1~$. By means of our power
counting rule (6), we can estimate that for leading contributions,
$T_1=O(\frac{E^4}{f^2_\pi \Lambda^2}f^{4-n}_\pi)$ and
$B=O(g^2 f^{4-n}_\pi)$~
\footnote{In the CLEWT, $f_\pi=246 GeV$
and the effective cut-off
$\Lambda \simeq 4 \pi f_\pi\simeq 3.1 TeV$.}.
Thus condition (4) requires
$\frac{M^2_W}{E^2}\ll \frac{1}{4}\frac{E^2}{\Lambda^2}$, or
$~(0.7 {\rm TeV}/E)^4 \ll 1~$.
So the probe is generally sensitive
when $E\geq 1$ TeV which is possible at the LHC.

In the CLEWT, the Lagrangian can be
written in the following form$^{6}$
$$
\begin{array}{ll}
{\cal L}_{eff} = {\cal L}_G + {\cal L}_F
+ {\cal L}^{(2)} + {\cal L}^{(2)\prime}
+ \displaystyle\sum_{n=1}^{14} {\cal L}_n&
= \displaystyle\sum_n \ell_n\displaystyle\frac{f_\pi~^{r_n}}{\Lambda^{a_n}}
{\cal O}_n(W_{\mu\nu},B_{\mu\nu},DU,U,f,\bar{f}) ,
\end{array}
\eqno(5)                          
$$
where ${\cal L}_G$,${\cal L}_F$ are the
kinetic  terms of the gauge fields
and fermions. The explicit formula for ${\cal L}_{eff}$ is
given in Ref.5$\sim$6, in
which ${\cal L}^{(2)}, {\cal L}^{(2)\prime},{\cal L}_{1\sim 11}$
are $CP$ conserving, and ${\cal L}_{12\sim 14}$  are $CP$
violating. Here, the dimensionless coefficients $\ell_n$'s
can be naturally regarded as of $O(1)$~$^7$.
In Ref.5, we developed the following power counting
rule in the CLEWT for the S-matrix element ~$T$
$$
\begin{array}{l}
T= c_T f_\pi^{D_T}\displaystyle
\left(\frac{f_\pi}{\Lambda}\right)^{N_{\cal O}}
\left(\frac{E}{f_\pi}\right)^{D_{E0}}
\left(\frac{E}{\Lambda}\right)^{D_{EL}}
\left(\frac{M_W}{E}\right)^{e_v} H(\ln E/\mu) ~~,\\[0.5cm]
N_{\cal O}=\displaystyle\sum_n a_n~,~~~~
D_{E0}=2+\displaystyle\sum_n {\cal V}_n(d_n+\frac{1}{2}f_n-2)~, ~~~~
D_{EL}=2L~,\\
\end{array}
\eqno(6)                                                  
$$
where the dimensionless coefficient $~c_T~$ contains
possible powers of gauge
couplings ($~g,g^\prime~$) and Yukawa
couplings ($~y_f~$) from the vertices
in $~T~$, which can be easily determined from
the vertices. $~H$ is a function
of $~\ln (E/\mu )~$ insensitive to $E$, where $\mu$ denotes the relevant
renormalization scale. $d_n$ is the
number of derivatives in the type-$n$
vertex, ${\cal V}_n$ is the number of type-n vertices in $T$,
$f_n=2i_F+e_F$ is the number of fermion fields.

\begin{table}[t]
\begin{center}
\footnotesize
{\bf Table 1.} Contributions of the model-dependent operators to the
$W^\pm$$W^\pm$$\rightarrow$$W^\pm$$W^\pm$ amplitudes
\vspace{0.8cm}

\renewcommand{\baselinestretch}{0.05}
\footnotesize
\begin{tabular}{||c||c|c|c|c|c||}
\hline\hline
& & & & &  \\
Operators
& $ T_1[4\pi] $
& $ T_1[3\pi,W_T] $
& $ T_1[2\pi,2W_T] $
& $ T_1[\pi,3W_T] $
& $ T_1[4W_T] $ \\
& & & & &  \\
\hline\hline
& & & & &  \\
$ {\cal L}^{(2)\prime} $
& $ \ell_0 ~\frac{E^2}{\Lambda^2} $
& $ \ell_0~g\frac{f_\pi E}{\Lambda^2} $
& $ \ell_0~g^2\frac{f_\pi^2}{\Lambda^2} $
& $ \ell_0~g^3\frac{f_\pi^3}{E \Lambda^2} $
& /  \\
& & & & &  \\
\hline
& & & & &  \\
$ {\cal L}_{1,13} $
&  /
& $ \ell_{1,13}~e^2g\frac{f_\pi E}{\Lambda^2} $
& $ \ell_{1,13}~e^4\frac{f_\pi^2}{\Lambda^2} $
& $ \ell_{1,13}~e^2g\frac{f_\pi E}{\Lambda^2} $
& $ \ell_{1,13}~e^2g^2\frac{f_\pi^2}{\Lambda^2} $  \\
& & & & &  \\
\hline
& & & & &  \\
$ {\cal L}_2 $
& $ \ell_2~e^2\frac{E^2}{\Lambda^2} $
& $ \ell_2~e^2g\frac{f_\pi E}{\Lambda^2} $
& $ \ell_2~e^2\frac{E^2}{\Lambda^2} $
& $ \ell_2~e^2g\frac{f_\pi E}{\Lambda^2} $
& $ \ell_2~e^2g^2\frac{f_\pi^2}{\Lambda^2} $ \\
& & & & &  \\
\hline
& & & & &  \\
$ {\cal L}_3 $
& $ \ell_3~g^2\frac{E^2}{\Lambda^2} $
& $ \ell_3~g\frac{E}{f_\pi}\frac{E^2}{\Lambda^2} $
& $ \ell_3~g^2\frac{E^2}{\Lambda^2} $
& $ \ell_3~g^3\frac{f_\pi E}{\Lambda^2} $
& $ \ell_3~g^4\frac{f_\pi^2}{\Lambda^2} $ \\
& & & & &  \\
\hline
& & & & &  \\
$ {\cal L}_{4,5} $
& $ \ell_{4,5}~\frac{E^2}{f_\pi^2}\frac{E^2}{\Lambda^2} $
& $ \ell_{4,5}~g\frac{E}{f_\pi}\frac{E^2}{\Lambda^2} $
& $ \ell_{4,5}~g^2\frac{E^2}{\Lambda^2} $
& $ \ell_{4,5}~g^3\frac{f_\pi E}{\Lambda^2} $
& $ \ell_{4,5}~g^4\frac{f_\pi^2}{\Lambda^2} $ \\
& & & & &  \\
\hline
& & & & & \\
$ {\cal L}_{6,7,10} $
& /
& /
& /
& /
& /  \\
& & & & &  \\
\hline
& & & & &  \\
$ {\cal L}_{8,14} $
& /
& $ \ell_{8,14}~g^3\frac{f_\pi E}{\Lambda^2} $
& $ \ell_{8,14}~g^2\frac{E^2}{\Lambda^2} $
& $ \ell_{8,14}~g^3\frac{f_\pi E}{\Lambda^2} $
& $ \ell_{8,14}~g^4\frac{f_\pi^2}{\Lambda^2} $ \\
& & & & & \\
\hline
& & & & & \\
$ {\cal L}_9 $
& $ \ell_9~g^2\frac{E^2}{\Lambda^2} $
& $ \ell_9~g\frac{E}{f_\pi}\frac{E^2}{\Lambda^2} $
& $ \ell_9~g^2\frac{E^2}{\Lambda^2} $
& $ \ell_9~g^3\frac{f_\pi E}{\Lambda^2} $
& $ \ell_9~g^4\frac{f_\pi^2}{\Lambda^2} $ \\
& & & & & \\
\hline
& & & & & \\
$ {\cal L}_{11,12} $
& /
& $ \ell_{11,12}~g\frac{E}{f_\pi}\frac{E^2}{\Lambda^2} $
& $ \ell_{11,12}~g^2\frac{E^2}{\Lambda^2} $
& $ \ell_{11,12}~g^3\frac{f_\pi E}{\Lambda^2} $
& $ \ell_{11,12}~g^4\frac{f_\pi^2}{\Lambda^2} $ \\
& & & & & \\
\hline\hline
\end{tabular}
\end{center}
\end{table}

\renewcommand{\baselinestretch}{1.3}
With this counting rule, we can
estimate the sensitivities  to probing specific
operators in (5) via various $W$-$W$
scattering amplitudes. In Table-1, we
list the results in the important
$W^\pm$$W^\pm$ channel as a typical example. We first
see that ${\cal L}_{6,7,10}$ do not contribute
to this channel. Table-1 then
shows that the $4W^\pm_L$ channel
can probe ${\cal L}_{4,5}$ most sensitively,
while the contributions of
${\cal L}^{(2)\prime},~{\cal L}_{2,3,9}$ to this channel
lose $E$-power dependence by a factor-$2$.
This channel cannot probe ${\cal L}_{1,8,11\sim 14}$.
${\cal L}_{1,8,11\sim 14}$ can only
be probed via channels with $W^\pm_T$('s),
among which $~{\cal L}_{11,12}~$ are most dominant though they
are still suppressed by a factor $gf_\pi /E$ relative to the
leading contributions to the $4W^\pm_L$ channel.
${\cal L}_{1,8,13,14}$ are generally suppressed by higher powers of
the factor $gf_\pi /E$ and are thus less sensitive.
For a more complete classification, see Table-3 in Ref.5.

We have further calculated the number of events
per $[100{\rm fb}^{-1}\cdot {\rm GeV}]$
at the LHC from our counting rule (6) combined with
the effective-$W$ approximation.
We have compared them with the corresponding available explicit
calculations in Ref.8 for a few typical examples. The
comparison shows that the deviations are reasonably within a
factor-$2$ which is of the same order as the uncertainty of the
effective-$W$ approximation$^9$. Therefore our power
counting rule does give correct semi-quantitative results and is
thus very useful and convenient for making a systematical analysis
for the sensitivities to probing
the SSB mechanism at the LHC and future
linear colliders. In the typical case with $\ell_n \sim O(1)$,
the number of the LHC events for the $W^+W^+$ channel are shown
in Fig.1. By comparing with the events from $B$, we see
that the probe of ${\cal L}_{4,5}$ are most sensitive, that of
${\cal L}_{3,9,11,12}$ are marginal,
and that of ${\cal L}^{(2)\prime},
{\cal L}_{1,2,8,13,14}$ are
insensitive. More of the details are given in Ref.5.

\vspace{0.50cm}
\noindent
{\bf Acknowledgements}~
H.J.H. is supported in part by the U.S. DOE under grant DEFG0592ER40709;
Y.P.K. is supported by the National
NSF of China and the FRF of Tsinghua
University; C.P.Y. is supported
in part by the NSF under grant PHY-9309902.

\newpage
\noindent
{\small {\bf Fig.1}  Sensitivities of operators ${\cal L}^{(2)\prime},
{\cal L}_{1\sim 14}$ whith $\ell_n\sim O(1)$, at the $14$ TeV LHC}.
\\ \vspace*{0.125in} \\ \hspace*{0.125in}
\psfig{file=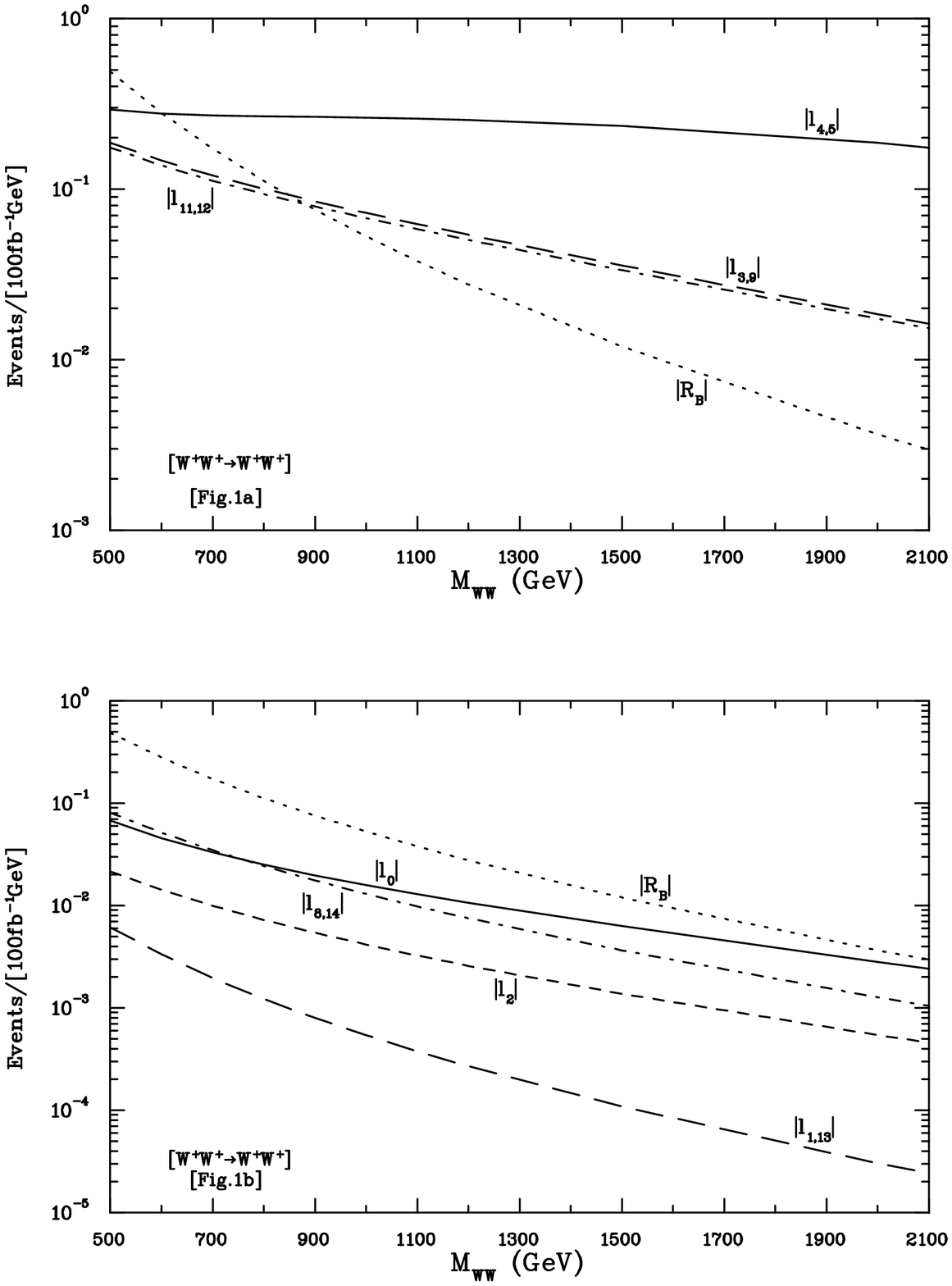,height=7.5in}

\end{document}